# Non thermal and purely electronic resistive transition in narrow gap Mott insulators

P. Stoliar*

*Laboratoire de Physique des Solides, CNRS UMR 8502, Université Paris Sud, Bât 510, 91405 Orsay, France and*

*ECyT, Universidad Nacional de San Martín, Irigoyen 3100, 1650 San Martín, Argentina*

M. Rozenberg

*Laboratoire de Physique des Solides, CNRS UMR 8502, Université Paris Sud, Bât 510, 91405 Orsay, France*

E. Janod, B. Corraze, J. Tranchant and L. Cario*

*Institut des Matériaux Jean Rouxel (IMN), Université de Nantes, CNRS, 2 rue de la Houssinière, BP32229, 44322 Nantes, France*

*corresponding authors: pstoliar@nanogune.eu, laurent.cario@cnrs-imn.fr

Mott insulator to metal transitions under electric field are currently the subject of numerous fundamental and applied studies. This puzzling effect, which involves non-trivial out-of-equilibrium effects in correlated systems, is indeed at play in the operation of a new class of electronic memories, the "Mott memories". However the combined electronic and thermal effects are difficult to disentangle in Mott insulators undergoing such transitions. We report here a comparison between the properties under electric field of a canonical Mott insulator and a model built on a realistic 2D resistor network able to capture both thermal effects and electronic transitions. This comparison made specifically on the family of narrow gap Mott insulators $AM_4Q_8$, (A = Ga or Ge; M=V, Nb or Ta, and Q = S or Se) unambiguously establishes that the resistive transition experimentally observed under electric field arises from a purely electronic mechanism.

**PACS** :
71.30.+h    Metal-insulator transitions and other electronic transitions
05.10.-a    Computational methods in statistical physics and nonlinear dynamics
71.27.+a    Strongly correlated electron systems; heavy fermions
64.60.ah    Percolation

The possibility of inducing an Insulator to Metal Transition (IMT) in correlated compounds by application of electrical fields $E$ [1-4] is currently receiving considerable attention due to the potentially new physics involved as well as for the possible applications in electronics [5,6]. On the fundamental side, a $E$-controlled IMT may result from emerging out-of-equilibrium mechanisms [7,8] going beyond the well established bandwidth- and filling-controlled IMT in Mott insulators.[9] On the applicative side, the new class of "Mott memories" based on this $E$-controlled IMT property, also referred to as resistive switching, [10] is a promising candidate for the next generation of electronic memories [11-13].

A central issue in this field is the identification of basic mechanisms leading to resistive transition in correlated insulators. Recent researches have shown that they can be sorted in three groups: thermal effects, electric-field-assisted ionic migration and electronic effects. In the first group, Joule heating can trigger a sharp resistive transition in materials undergoing an IMT driven by temperature. Archetypical examples are the celebrated Mott insulators $VO_2$ [9,17,18,20] and $NbO_2$ [19]. In these compounds, the strong drop of electrical resistivity at the IMT temperature is accompanied by a change in lattice symmetry. In a second group, migration of ions (for example oxygen atoms) in a correlated material can lead to a local non-stoichiometry and therefore to a local filling-controlled IMT which destabilizes the Mott insulating state [6,14]. In such systems, an electric-field-assisted displacement of ions occurs at the interfaces [15] and leads to bipolar resistive transition. The third group comprises purely electronic mechanisms, such as the theoretically-predicted destabilization of the correlated state in strong electric field [7], also referred to as Zener dielectric breakdown of Mott insulators [8]. In such a mechanism, the IMT occurs without changes in the lattice symmetry [16]. Recently, an additional purely electronic mechanism of resistive transition has emerged [2] from studies of the prototypical family of Mott insulators $AM_4Q_8$ (A=Ga, Ge; M=V, Nb, Ta; Q=S, Se, Te) [28]. In these compounds, a resistive transition occurs above a moderate threshold electric field $E_{th}$ in the 1-10 kV/cm range. The observation of an $E_{th}$ vs. $E_{gap}$ dependence very similar to the classical semiconductors leads to propose that an electronic avalanche is the driving force of the resistive transition in these Mott insulators [2]. The other so far identified mechanisms of resistive transition seems indeed unsuitable for the $AM_4Q_8$ compounds. The unipolar and non-interfacial character of the resistive transition easily excludes mechanisms related to ionic migration. Conversely, thermal effects are more tricky to rule out. On the one hand, $AM_4Q_8$ compounds do not display any insulator to metal transition driven by temperature as observed in $VO_2$. On the other hand, $AM_4Q_8$ compounds display a strongly temperature-dependent resistivity associated with a Mott-Hubbard gap of 0.1-0.3 eV. In this context, the possibility that the observed resistive transition results from the sudden formation of a hot and highly conductive filament induced by Joule heating [32] can not be dismissed out of hand.

In this work, we show that the resistive transition can not arise from Joule heating effects in $AM_4Q_8$ compounds and that these narrow gap Mott insulators are prototypical system for the study of purely electronic resistive switching in correlated systems. We compare key experimental observations with calculations performed for both the hot filament and an electric-field-driven collapse of the Mott state scenarios. The experimental results are in clear disagreement with the thermal model characteristics, whereas they present all the features predicted by a purely electronic model. This strongly suggests that the resistive switching in these systems is primarily driven by electronic rather than thermal effects.

The phenomenology of the resistive switching in the $AM_4Q_8$ compounds was comprehensively reported in previous publications [29-31]. The typical experiments relevant to this study are performed on $AM_4Q_8$ single crystals with two metal contacts on opposite faces. Initially the system is in Mott insulating state, with a resistance $R_0$. A voltage pulse $V_{pulse}$ is applied on a dipole composed of the sample in series with a limiting resistor $R_L \approx$ $R_0/10$. Above a threshold field of a few kV/cm the sample undergoes a sharp decrease of its resistance during the pulse (the electric field is defined as the voltage on the sample, $V_{sample}$, divided by the distance between the electrodes, $d$). This effect is called *volatile* resistive switching since the resistance of the sample returns to its initial value after the pulse. For even higher electric fields (about 3 to 5 times the threshold field) the resistive switching becomes *non-volatile*, meaning that the sample remains in low resistance state after the electric pulse. In the rest of this work we will only focus on the volatile transition of the $AM_4Q_8$-based devices as both the volatile and the non-volatile resistive switching derive from the same initial physical mechanism. [30]

Fig. 1(a) displays the change in resistance of the $GaTa_4Se_8$ crystal ($R/R_0$) during the application of voltage pulses. Resistive switching is observed for voltages that exceed the threshold voltage $V_{th} \approx 10$ V corresponding to a threshold field $E_{th} \approx 2.5$ kV/cm ($E_{th} = V_{th}/d$). The time $t_{delay}$, at which the sharp resistive switching occurs, decreases when the voltage pulse increases (see Fig. 1(b)). The sample voltage $V_{sample}$ after the resistive switching is always the same and is equal to $V_{th} \approx 10$ V whatever the voltage applied to the sample before the resistive switching. Consequently, the current flowing through the sample $I$ after the switching varies in a large extent. Fig. 1(c) shows the current-voltage characteristic, which is highly non-ohmic and appears as an almost vertical line in the "transited" state. As we shall see, this voltage stabilization is a key feature in our study.

In order to address the question whether the creation of a hot filament by Joule heating could account for the resistive switching in $AM_4Q_8$ compounds, an experimental approach could be envisioned. For example, a "tour de force" experiment was recently performed, by using the fluorescence spectra of rare-earth doped micron sized particles as local temperature sensors at the surface of a $VO_2$ thin layers [20]. However, addressing the question of a hot filament is much more demanding in our case, since it also requires both a submicron spatial resolution in the bulk of the AM$_4$Q$_8$ single crystals, and also a microsecond time resolution. As such experimental approaches are to our knowledge out of reach, we have thus adopted a strategy based on thermal modeling. We have first tested the hypothesis of the formation of hot conductive filaments within a thermal model based on the activated dependence of the resistivity in these compounds. We thus adopted the two-dimensional resistor network model shown in Fig. 2(a) to simulate the resistance and local thermal variations of our sample. Each cell (grayed box in the figure) is named $C_{y,x}$ ; $x$ = 1 ... 64, y = 1 ... 25, *i.e.* we have 25 lines of cells between the electrodes. Each cell consists of four resistors connected to a single node in cross shape. In addition, for the present thermal model calculation, the cell has four thermal conductances $\kappa$ and a heat capacity $c$. Periodic boundary conditions are assumed in the x-direction. The four resistors of a given cell have the same value, which depend on the temperature of the cell, $T(x,y)$, following the activated law

$$r(x,y) = r_0 \exp\left(\frac{E_A}{k_B T(x,y)}\right),$$  (1)

where $E_A$ is the activation energy for the electrical transport, $k_B$ the Boltzmann constant and $r_0$ an arbitrary pre-factor [33,34]. We chose $E_A$ and $r_0$ so that $r(x,y)$ fits accurately the temperature dependence of the experimental resistivity. The simulation time is discretized in time-steps. At each time-step the 2D resistor network is solved for the applied external voltage by standard nodal analysis. A temperature $T_A$ is initially set for all the sites. To account for defects present in the bulk material and the imperfections of the electrodes, we set the resistance of some few cells (< 1%) with a fixed $r_{defect}$, $10^4$ times lower than $r(x,y)$ at $T_A$. From the calculated local currents, the change in temperature of each site is computed and updated. This is done using the heat equation, expanding $\nabla^2 T(x,y)$ by means of the discrete Laplacian approximation:

$$\frac{\partial T(x,y)}{\partial t} c = p(x,y) - \kappa \cdot \left( 4T(x,y) - \sum_{i}^{\text{1st neighbors}} T_i \right), \qquad (2)$$

where $c$ is the heat capacity of each cell, $p(x,y)$ the electric power generated in the cell and $\kappa$ the thermal conductance between neighboring sites (same for all the network). As boundary condition, the cells at the top and at the bottom exchange heat with the 3 neighbors and a virtual cell with a fixed temperature $T_A$, which represents the electrodes. Based on this new temperature profile, the matrix $r(x,y)$ is updated, and then the simulation moves on to the next time-step [35].

Typical results of the simulation are presented in Fig. 1(d) – (f). A gradual decrease of the resistance is observed for all applied voltages due to Joule heating, eventually arriving to a stable low resistance condition. Also, sequences showing the evolution of the temperature profile at different simulation time are presented in Fig. 2(b). We observe the formation of a hot percolative path bridging the electrodes. Due to the activated temperature dependence, the resistivity of this hot filamentary path decreases, leading to a reduction of the sample resistance. This resistance drop can reach several orders of magnitude for large $V_{pulse}$. As observed in the experiments, $t_{delay}$ decreases as $V_{pulse}$ increases. However, a quick comparison with the experimental data reveals that the Joule heating model fails to reproduce the $R_{sample}(t)$ curve and the I-V characteristics. Unlike the experiments, simulations never present a sharp drop of the resistance at a ($V_{pulse}$-dependent) delay-time measured from the beginning of the application of the pulse. Moreover, further key discrepancies with experiments are that the model fails to predict the existence of a defined threshold voltage and the voltage regulation effect that is observed after the transition. In fact, all applied $V_{sample}$ do produce similar transients, but the asymptotic $V_{sample}$ does not reach the same final value after switching. Instead the final $V_{sample}$ value decreases with increasing $V_{pulse}$. As a consequence, the I-V characteristic in the stationary regime (red circles in Fig. 1(f)) displays a typical negative differential resistance (NDR, [32]), which is in strong contrast with experiments. This NDR regime is a clear indication of the generation of the high-current/low-resistivity filaments visualized in Fig. 2(b). The almost vertical I-V characteristic found in the $AM_4Q_8$-based devices reflects a zero differential resistance that evidences a mechanism of a different nature (see red circles in Fig. 1(c)).

At this step it is clear that the resistive switching observed in the $AM_4Q_8$-based devices does not arise from a purely Joule heating process. We have therefore implemented a purely electronic model as introduced in Ref. [36]. This model, illustrated in Fig. 3(a), assumes that the active layer is composed of cells that may be in either one of only two states; a low-resistivity correlated metal state (CM) and a high-resistivity Mott insulating state (MI). For each cell, an energy barrier $E_B$ must be overcome to switch from the MI to the CM state. The cells are modeled by four resistors with value $r_{MI}$ when it is in MI state or $r_{CM}$ when it is in CM state, with $r_{CM} < r_{MI}$. The electronic configuration of the Mott insulating state is more stable (the experimental system is in fact a single crystal Mott insulator), so that in thermal equilibrium the device is initially assumed to be in high-resistance state (HRS). Each cell should represent a small portion of the sample, such that the local resistivity may have a well-defined value, that is, of at least a few tenths of nanometers.

We assume that the electronic configuration of a cell in Mott insulating state can be destabilized by the effect of a local electric field, $\varepsilon$. It effectively reduces the $E_B$ barrier, resulting in a probability

$$P_{CM}(x,y) = \exp\left(-\frac{E_B - q|\varepsilon(x,y)|}{k_B T(x,y)}\right) \tag{3}$$

for the local "collapse" of the Mott insulating state towards the CM one. We initially keep $T(x,y)$ at a fixed uniform and constant value during the simulations (we shall later relax this constraint). When a cell commutes to the highly conductive CM state, its internal electric field becomes negligible ($\varepsilon \approx 0$ inside a metallic region), so that the cell will have a probability

$$P_{MI}(x,y) = \exp\left(-\frac{E_B - E_M}{k_B T(x,y)}\right) \tag{4}$$

for switching back to the Mott insulating state.

Fig. 3(a) presents the array of cells connected to the pulse source $V$ through the limiting resistor $R_L$. We simulated the time evolution of this system similarly as before. The sole difference is that from the computed voltage drops at each cell and by means of Eq. (3) and Eq. (4), we evaluate whether a cell must change its state or not, and the value of the resistors of the network are updated accordingly [37].

Fig. 1(g) – (i) show the typical results of the model simulations. In this case, the computed curves qualitatively reproduce the experiment, that is, the delayed sudden drop of the resistance above a voltage threshold (Fig. 1(g)), the scaling of $t_{delay}$ with the voltage (Fig. 1(h)) and the voltage regulation effect after switching (Fig. 1(i)).

The simulations with this purely electronic model predict two different regimes depending on the applied voltage; a low voltage regime where it remains in the HRS, and a high voltage regime where it undergoes a transition to a Low Resistance State. In the low voltage regime, the stationary state corresponds to a diluted condition, *i.e.* with the system largely in a homogeneous MI state, and just a few isolated CM cells. There, equilibrium is reached, when the rate of MI cells switching to CM state equals the rate of CM cells switching back to the MI state (cf. Eqs. 3 and 4). In this situation, the electric field remains essentially homogeneous across the system. Instead, in the high voltage regime, the density of CM cells steadily increases in time. At this higher voltage, there is a significant probability for the formation of metallic clusters, which create spatial inhomogeneities in the electric field. Eventually, a high local field region will be the seed of a "runaway" event, which results in the sudden formation of a metallic filament that connects the electrodes, as illustrated in Fig. 3(b). The two regimes that we just describe characterize the existence of a threshold voltage, as observed in the experiments.

While it is now clear that the electronic model captures the key features of the resistive switching, it is also an experimental fact that the resistivity of these narrow-band insulators is very sensitive to local changes in temperature. Therefore, we shall finally combine the electronic (Fig. 3) and the thermal models in order to study the interplay between the IMT and the Joule heating. Now, the temperature at each cell is calculated as described before, and the switching probabilities $P_{CM}$ and $P_{MI}$ (Eq. (3) and Eq. (4)) depend on the computed local temperature. Additionally, the resistances in the Mott insulating state are now considered temperature dependent, following Eq. (1).

Fig. 4 presents the results of this combined model. We observe that the increase in temperature before switching is minimal. The switching behavior is set off by the mechanism associated to the electronic model, *i.e.* it is dominated by the electric-field-driven IMT. After switching, the temperature of the filament first rapidly increases and, after a voltage dependent time-scale, reaches a stable state with a higher conductance. Therefore both behaviors are present, but operate at different time-scales. At short times we observe voltage regulation after switching. At longer times this stabilization is lost due to filament Joule heating that might ultimately lead to sample destruction.

In conclusion, we compared resistive switching experimental results in $AM_4Q_8$ compounds to the predictions of two models that incorporate different physical mechanisms that have been evoked in the literature as likely candidates to capture the insulator to metal transition in correlated Mott insulators under strong electric pulsing: a purely thermal and a purely electronic model. Our results showed that the former might capture the lowering of the resistance by the formation of a highly conductive hot filament, but misses key experimental observations such as the sharpness of the resistive transition and the precise voltage regulation after switching. On the other hand, our results demonstrate that the latter model, based on the electric collapse of the Mott state, successfully reproduces all the qualitative experimental features. Finally, taking into account thermal effects in the electronic model leaves this main conclusion unchanged: the driving force of the resistive transition in $AM_4Q_8$ is an electronic effect. We note that the "switched" state in $AM_4Q_8$, consisting basically in percolating conducting filaments embedded in a Mott insulating matrix, strongly differs from the one in "Mott compounds" exhibiting a temperature driven IMT. In compounds such as $VO_2$, the entire sample can switch to a bulk and homogeneous metallic state for large enough Joule heating. More generally, our results and conclusions are likely of a broader reach. In fact, they may be relevant also to other narrow gap Mott insulators, including the canonical system Cr-doped $V_2O_3$ and $Ni(S,Se)_2$ where similar effects were recently uncovered [36]. Important theoretical challenges remain ahead, in particular a full many-body out-of-equilibrium treatment of the Mott transition, which remains a formidable problem. Therefore, numerical studies, such as the present one, are urgently needed to provide useful insights and guidance to the current fast progress of the emergent field of correlated electronics.

## Acknowledgments

This work was supported by the French Agence Nationale de la Recherche through the funding of the "NanoMott" (ANR-09-Blan-0154-01), "Gold-RRAM" (ANR-12-BS07-0032-03) and "Mott-RAM" (ANR-2011-EMMA-016-01) projects. The authors acknowledge S. Salmon for sample preparation, M.-P. Besland, D. Roditchev, T. Cren and V. Ta Phuoc for useful discussions.

# Figures

*FIG. 1*

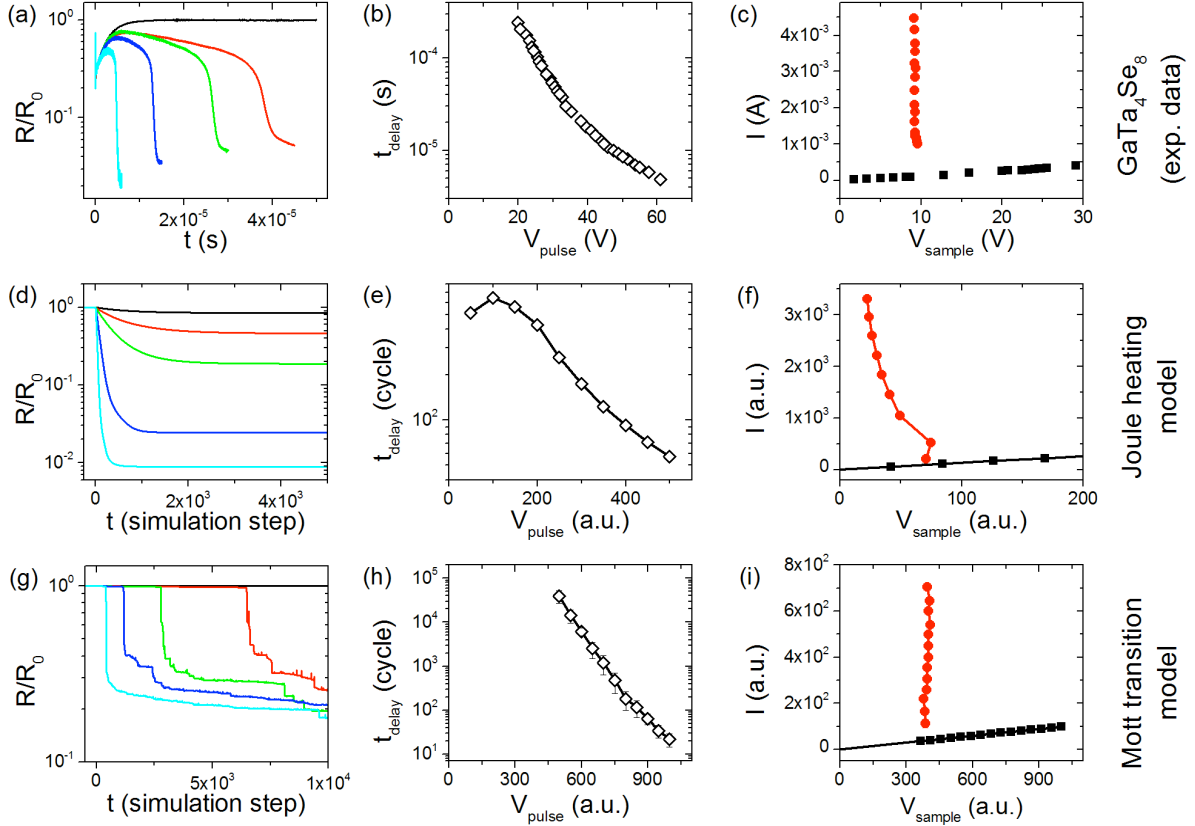

Comparison between the experimental data (a-c) and the two studied models, the Joule heating model (d-f) and the Mott transition model (g-i).[36] The measurements were conducted in a GaTa$_4$Se$_8$-based device at 77 K. For each case we present the variation of the resistance of the device in time for increasing $V_{pulse}$, the delay elapsed between the application of $V_{pulse}$ and the switching (if it occurs) and the I-V characteristic before and after the switching. The curves presented in (a) correspond to $V_{pulse}$ values of 6, 40, 44, 56 and 86V. The duration of pulses was limited to 50, 45, 30, 15 and 6 μs in order to avoid overheating of the sample [38]. For the thermal model, the $t_{delay}$ (e) is considered when the decrease of resistance is half of the final value and (f) plots the current and the voltage at the beginning of the simulations and after $10^5$ cycles, where the system was always in stationary regime. The curves in (d) correspond to $V_{pulse}$ values of 50, 100, 150, 300 and 500 a.u. and the curves presented in (g) to $V_{pulse}$ values of 400, 600, 650, 700 and 750 a.u.

*FIG. 2*

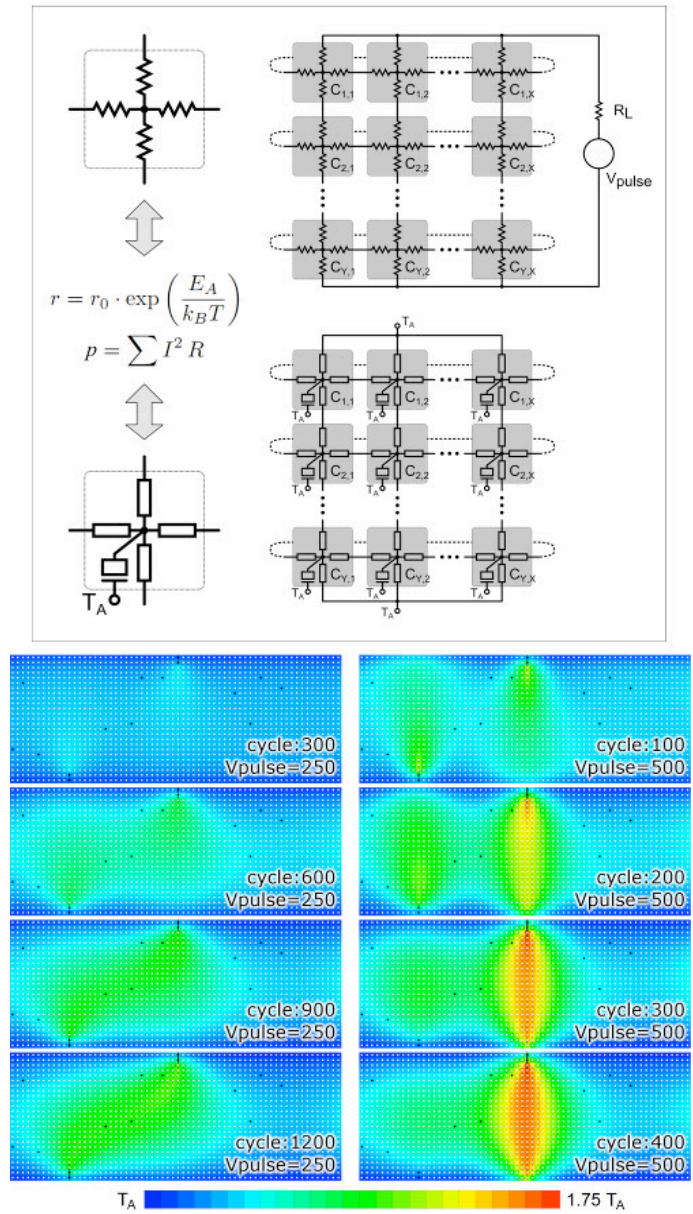

Concept of the Joule heating model (top) and simulated evolution of the temperature profile inside the device (bottom). The color scale represents local increase of the temperature above the initial $T_A$. The cells with black dots are considered as highly conductive defects. In all the figures throughout this work, voltage is applied between the electrodes located at the top and at the bottom of the resistor network.

*FIG. 3*

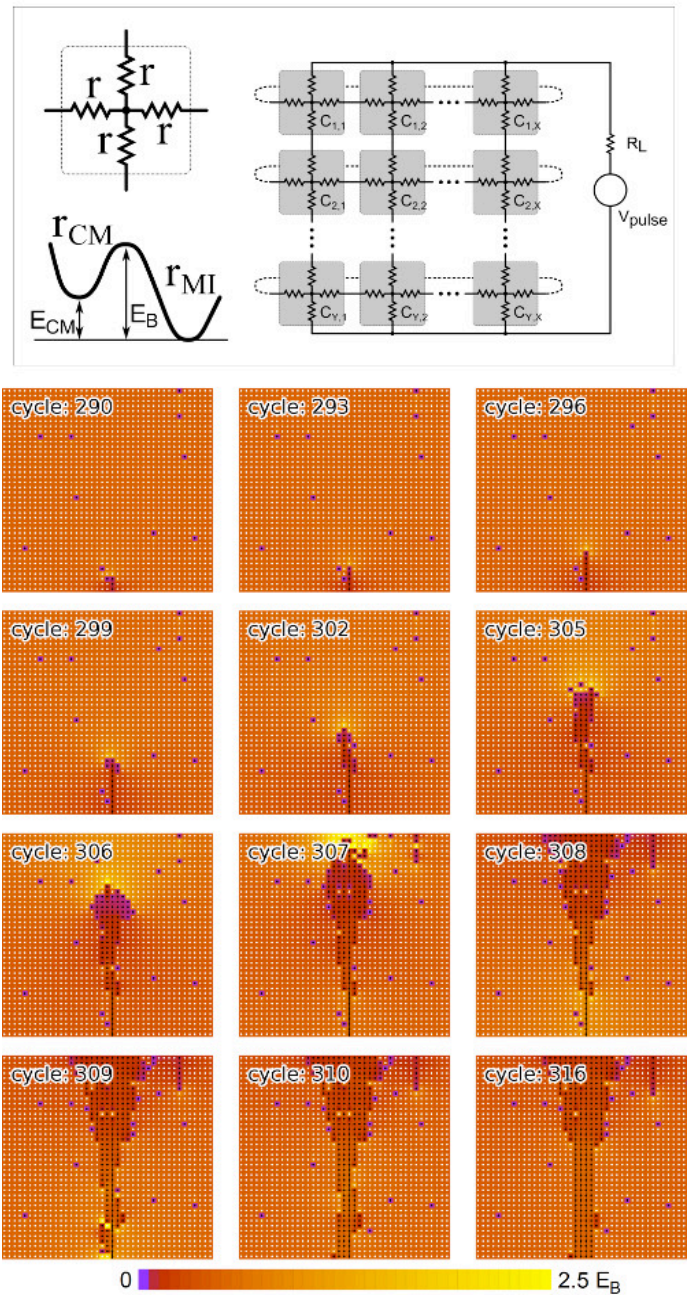

Concept of the Mott transition model (top) and evolution of the electric field profile in a region of the device (bottom). The color scale represents the local electric field *qε*. The state of each cell is represented by a white or black dot, corresponding to insulating and metallic cells respectively.

*FIG. 4*

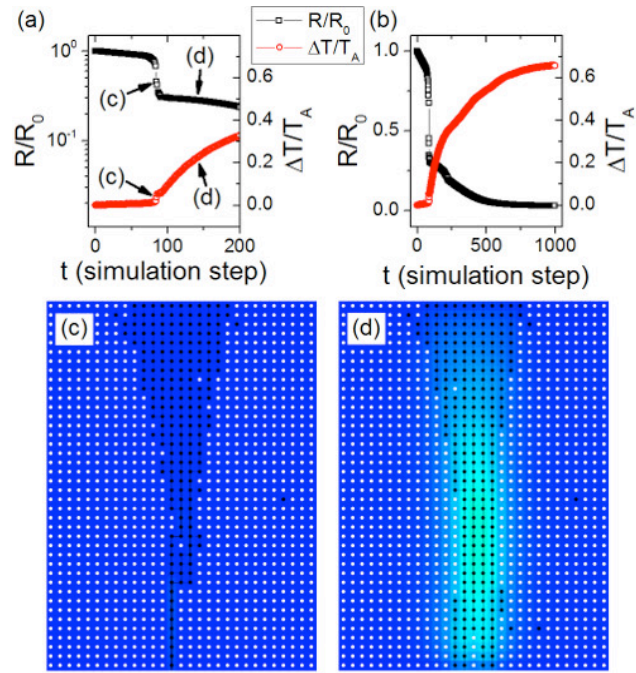

Output of the combined (electronic + thermal) model. Panels (a) and (b) present the time evolution of the resistance $R$ and the increase of the temperature in the hottest point $\Delta T$ at different time scales. At short time scale (a) the electronic switch is evident, whereas the progressive increase of the temperature is observed at longer times (b). The temperature profiles (c) and (d), with the same color scale than Fig. 2, show that the relevant Joule heating starts after the electronic switch.

# References

[1] Y. Taguchi, T. Matsumoto, and Y. Tokura, Phys. Rev. B **62**, 7015 (2000).

[2] V. Guiot, L. Cario, E. Janod, B. Corraze, V. T. Phuoc, M. Rozenberg, P. Stoliar, T. Cren, and D. Roditchev, Nature Communications **4**, 1722 (2013).

[3] F. Sabeth, T. Iimori, and N. Ohta, J. Am. Chem. Soc. **134**, 6984 (2012).

[4] H-S Lee, S-G Choi, H-H Park and M. J. Rozenberg, Scientific reports **3**, 17 (2013).

[5] H. Takagi and H. Y. Hwang, Science **327**, 1601 (2010).

[6] H. Y. Hwang, Y. Iwasa, M. Kawasaki, B. Keimer, N. Nagaosa, and Y. Tokura, Nat Mater **11**, 103 (2012).

[7] F. Heidrich-Meisner, I. Gonzalez, K. A. Al-Hassanieh, A. E. Feiguin, M. J. Rozenberg, and E. Dagotto, Phys. Rev. B **82**, 205110 (2010).

[8] T. Oka, Phys. Rev. B **86**, 075148 (2012). T. Oka and H. Aoki, Phys. Rev. B **81**, 033103 (2010).

[9] M. Imada, A. Fujimori and Y. Tokura, Rev. Modern Phys. **70**, 1039 (1998).

[10] M. Rozenberg, Scholarpedia **6**, 11414 (2011).

[11] A. Sawa, Materials Today **11**, 28-36 (2008).

[12] R. Waser and M. Aono, Nature Materials **6**, 833 - 840 (2007).

[13] International Technological Roadmap for Semiconductor (ITRS). Available at www.itrs.net

[14] Y.B. Nian, J. Strozier, N.J. Wu, X. Chen and A. Ignatiev, Phys. Rev. Lett. **98**, 146403 (2007).

[15] M. J. Rozenberg, I. H. Inoue, and M. J. Sánchez, Phys. Rev. Lett. **92**, 178302 (2004).

[16] D. B. McWhan and J. P. Remeika, Phys. Rev. B **2**, 3734 (1970).

[17] M. Qazilbash, M. Brehm, B. Chae, P. Ho, G. Andreev, B. Kim, S. Yun, A. Balatsky, M. Maple, F. Keilmann, et al., Science **318**, 1750 (2007).

[18] A. Sharoni, J. Ramirez, and I. Schuller, Phys. Rev. Lett. **101**, 26404 (2008).

[19] M. D. Pickett and R. S. Williams, Nanotechnology **23**, 215202 (2012).

[20] A. Zimmers, L. Aigouy, M. Mortier, A. Sharoni, S. Wang, K. West, J. Ramirez, and I. Schuller, Phys. Rev. Lett. **110**, 056601 (2013).

[21] T. Driscoll, H. Kim, B. Chae, M. Di Ventra, and D. Basov, App. Phys. Lett. **95**, 043503 (2009).

[22] J. Kim, C. Ko, A. Frenzel, S. Ramanathan, and J. Hoffman, App. Phys. Lett. **96**, 213106 (2010).

[23] Y. Pershin and M. Di Ventra, Advances in Physics **60**, 145 (2011).

[24] H. Kim, B. Chae, D. Youn, S. Maeng, G. Kim, K. Kang, and Y. Lim, New Journal of Physics **6**, 52 (2004).

[25] G. Stefanovich, A. Pergament, and D. Stefanovich, J. Phys.: Condens. Matter **12**, 8837 (2000).

[26] T. Driscoll, J. Quinn, M. Di Ventra, D. Basov, G. Seo, Y. Lee, H. Kim, and D. Smith, Phys. Rev. B **86**, 094203 (2012).

[27] $VO_2$ is a typical example of Mott insulator for which the mechanism of resistive switching is under debate. This compound undergoes an IMT just above room temperature that was ascribed by some authors to the Mott physics [9,17,18]. Application of a DC voltage of enough amplitude on a $VO_2$ sample, at a temperature lower than the IMT, lead to a volatile resistive switching that was recently ascribed to the crossing of the IMT transition due to self-Joule heating [20]. Non-volatile RS may be also observed in this compound resulting from the hysteretic character of the first-order temperature-driven IMT. In this case, after an electric pulse, the persistence of metallic nano-domains within the insulating matrix explains the lowered value of the resistance [21-23]. Most works performed so far in DC mode on $VO_2$ suggest that temperature driven IMT and thermal joule heating are key ingredients of the mechanism of this resistive switching [20,21]. But, voltage pulse experiments [24] or delay time measurements [25] as well as recent modeling [26] indicate that an electric field driven electronic transition rather than Joule heating could trigger the RS in $VO_2$. Some questions remain therefore open concerning this resistive switching as it is not always easy to disentangle between electric-field and thermal effects especially in Mott insulators with T-driven IMT like $VO_2$.

[28] V. Ta Phuoc, C. Vaju, B. Corraze, R. Sopracase, A. Perucchi, C. Marini, P. Postorino, M. Chligui, S. Lupi, E. Janod and L. Cario, Phys. Rev. Lett. **110**, 037401 (2013).

[29] (a) C. Vaju, L. Cario, B. Corraze, E. Janod, V. Dubost, T. Cren, D. Roditchev, D. Braithwaite, and O. Chauvet, Adv. Mater. **20**, 2760 (2008). (b) V. Dubost, T. Cren, C. Vaju, L. Cario, B. Corraze, E. Janod, F. Debontridder, and D. Roditchev, Advanced Functional Materials **19**, 2800 (2009). (c) V. Dubost, T. Cren, C. Vaju, L. Cario, B. Corraze, E. Janod, F. Debontridder, and D. Roditchev, Nano Lett. **13**, 3648 (2013).

[30] (a) L. Cario, C. Vaju, B. Corraze, V. Guiot, and E. Janod, Adv. Mater. **22**, 5193 (2010). (b) B. Corraze, E. Janod, L. Cario, P. Moreau, L. Lajaunie, P. Stoliar, V. Guiot, V. Dubost, J. Tranchant, S. Salmon, M.-P. Besland, V. T. Phuoc, T. Cren, D. Roditchev, N. Stéphant, D. Troadec, and M. Rozenberg, Eur. Phys. J. Spec. Top. **222**, 1046 (2013).

[31] (a) E. Souchier, L. Cario, B. Corraze, P. Moreau, P. Mazoyer, C. Estournes, R. Retoux, E. Janod, and M.-P. Besland, Phys. Status Solidi RRL **5**, 53 (2011). (b) J. Tranchant, E. Janod, L. Cario, B. Corraze, E. Souchier, J.-L. Leclercq, P. Cremillieu, P. Moreau, and M.-P. Besland, Thin Solid Films **533**, 61 (2013). (c) E. Souchier, M.-P. Besland, J. Tranchant, B. Corraze, P. Moreau, R. Retoux, C. Estournès, P. Mazoyer, L. Cario, and E. Janod, Thin Solid Films **533**, 54 (2013).

[32] B. K. Ridley, Proceedings of the Physical Society **82**, 954 (1963).

[33] C. Vaju, L. Cario, B. Corraze, E. Janod, V. Dubost, T. Cren, D. Roditchev, D. Braithwaite, and O. Chauvet, Microelectronic Engineering **85**, 2430 (2008).

[34] V. Guiot, E. Janod, B. Corraze, and L. Cario, Chem. Mater. **23**, 2611 (2011).

[35] In this work we set $T_{init} = 1$, $E_A/k_B = 18.42$, $r_0 = 10^{-8}$ and $R_L = 0.1443$ (matching the experimental conditions). $k = 10^4$ and $c = 10^5$ where set to have a maximum transient time of $10^5$ program cycles.

[36] P. Stoliar, L. Cario, E. Janod, B. Corraze, C. Guillot-Deudon, S. Salmon-Bourmand, V. Guiot, J. Tranchant, M. Rozenberg, Adv. Mater. **25**, 3222 (2013).

[37] For the numerical calculations we used the following parameters: $E_B = 20\ k_BT$, $E_M = 10\ k_BT$, $r_M = 0.3\ R_L$, $r_I = 16.41\ R_L$ and $q = l\ k_BT$, where $l$ represents the distance between two cells. The number of cells is 128 x 40.

[38] In the experiments reported here, an Agilent 8114A High Power Pulse Generator generated the voltage pulses, which is capable of nanosecond-range rise times. Nevertheless, the effective rise time of the sample voltage is in the order of 3-4 μs, mainly due to the time constant defined by straight capacitance of the wiring and the load and the device resistances.